\documentclass[12pt]{article}
\usepackage{amsmath}
\usepackage{amsmath,amssymb}

\begin{document}

\numberwithin{equation}{section}

\title{Finite amplitude inhomogeneous waves in Mooney-Rivlin
viscoelastic solids.}
\author{Michel Destrade, Giuseppe Saccomandi}
\date{2004}
\maketitle
\bigskip


\begin{abstract}

New exact solutions are exhibited within the framework of
finite viscoelasticity.
More precisely, the solutions correspond to finite-amplitude,
transverse, linearly-polarized,
inhomogeneous motions superposed upon a finite homogeneous static
deformation.
The viscoelastic body is composed of a Mooney-Rivlin viscoelastic
solid, whose constitutive equation consists in the sum of an elastic
part (Mooney-Rivlin hyperelastic model) and a viscous part (Newtonian
viscous fluid model).
The analysis shows that the results are similar to those obtained
for the purely elastic case; inter alia, the normals to the planes
of constant phase and to the planes of constant amplitude must be
orthogonal and conjugate with respect to the $\mathbf{B}$-ellipsoid,
where $\mathbf{B}$ is the left Cauchy-Green strain tensor associated
with the initial large static deformation.
However, when the constitutive equation is specialized either to the
case of a neo-Hookean viscoelastic solid or to the case of a Newtonian
viscous fluid, a greater variety of solutions arises, with no
counterpart in the purely elastic case.
These solutions include travelling inhomogeneous finite-amplitude
damped waves and standing damped waves.

\end{abstract}

\newpage

%


\section{Introduction}


Hayes and Saccomandi \cite{HS1} initiated the study of finite amplitude
waves in a simple class of viscoelastic materials of the differential type
with a special constitutive equation for the Cauchy stress tensor. For these
materials, denoted as \emph{Mooney-Rivlin viscoelastic materials}, the
Cauchy stress tensor is split into an elastic part, which coincides with the
Cauchy stress tensor for a Mooney-Rivlin elastic material, and a dissipative
part, which coincides with the Cauchy stress tensor for a Newtonian fluid
(linear in the stretching tensor). Coleman and Noll \cite{Col1} were the
first to investigate the possibility of splitting the Cauchy stress into an
elastic part and a dissipative part, and Fosdick and collaborators (Fosdick
and Yu \cite{F2}, Fosdick, Ketema, and Yu \cite{F4}) also considered
Mooney-Rivlin viscoelastic materials.

With respect to wave propagation, Hayes and Saccomandi \cite{HS1} obtained
all homogeneous plane waves that may propagate in Mooney-Rivlin viscoelastic
materials maintained in a static state of pure homogeneous deformation. Then
they showed \cite{HS2} that principal transverse homogeneous waves may
propagate when superimposed on a special class of pseudo-plane inhomogeneous
steady motions. Finally, they considered antiplane shear waves superimposed
on a static biaxiall stretch \cite{HS3}.

These results \cite{HS1,HS2,HS3} provide a formidable corpus of exact
solutions that may be useful as benchmarks for more complicated problems.
Moreover, these solutions allow for a better understanding of the complex
subject of nonlinear viscoelasticity \cite{Ogden}. The aim of this note is
to extend these results to the case of \textit{inhomogeneous} plane waves
that is, plane waves for which the planes of constant phase are not parallel
to the planes of constant amplitude. Here we show that finite amplitude,
linearly polarized, transverse, inhomogeneous plane waves superimposed on a
generic homogeneous deformation are rare because two strict conditions must
be met: first, the planes of constant phase must be orthogonal to the planes
of constant amplitude; second, the equations of motion lead to a set of two
differential equations to be satisfied simultaneously. However, when the
constitutive equation is specialized from Mooney-Rivlin to neo-Hookean
viscoelastic materials and further, to viscous Newtonian fluids, then these
strict conditions disappear and many more exact solutions are obtained,
including finite-amplitude inhomogeneous damped plane waves. A remarkable
feature of the solutions is that although the static strains are finite, the
viscoelasticity theory nonlinear, and the waves of a generic (separable)
finite amplitude inhomogeneous travelling type, the governing equations
nevertheless eventually reduce to a set of \textit{linear} differential
equations which are solved exactly. No small parameters nor asymptotic
expansions are required, in contrast with most studies of finite-amplitude
waves in solids (see Norris \cite{Norr98} for a review). The only
restrictions lie with the specificity of the constitutive equation and with
the form of the finite-amplitude inhomogeneous plane waves. This work
follows the path laid out by Hadamard \cite{Hada03} for small-amplitude
homogeneous plane waves in finitely deformed compressible elastic materials;
by John \cite{John66} and by Currie and Hayes \cite{CuHa69} for
finite-amplitude homogeneous waves in deformed elastic compressible
``Hadamard'' materials and incompressible Mooney-Rivlin materials,
respectively; and by Destrade \cite{D} for finite-amplitude inhomogeneous
waves in deformed elastic Mooney-Rivlin materials.

The plan of the paper is the following. In the next Section we introduce the
basic equations governing the constitutive model of a Mooney-Rivlin
viscoelastic material, and the propagation of finite amplitude plane
inhomogeneous waves superposed on a large static pure homogeneous
deformation. In Section 3, the equations of motion are solved and we find
the conditions for the directions of polarization, of propagation, and of
attenuation under which the waves may propagate. Sections 4 and 5 are
devoted to the special cases of neo-Hookean viscoelastic materials and of
Newtonian viscous fluids.


\section{Preliminaries}



\subsection{Constitutive equations}


The elastic part of the model is characterized by the Mooney-Rivlin
strain-energy density $W$ which, measured per unit volume in the undeformed
state, is given by
\begin{equation}  \label{01}
2W = C(\text{I} - 3) + D(\text{II} - 3) ,
\end{equation}
where the constants $C$, $D$ satisfy $C \ge 0$, $D >0$ or $C>0$, $D \ge 0$
\cite{BoHa95}. The sum $C+D$ is the infinitesimal shear modulus and $\text{I}
$, $\text{II}$ denote the first and second principal invariants of the left
Cauchy-Green strain tensor $\mathbf{B} = \mathbf{FF}^\text{T}$:
\begin{equation}  \label{02}
\text{I} = \text{tr }\mathbf{B}, \quad 2\text{II} = (\text{tr }\mathbf{B})^2
- \text{tr}(\mathbf{B}^2).
\end{equation}
The components of the gradient of deformation $\mathbf{F}$ are
\begin{equation}  \label{03}
F_{iA}= \frac{\partial x_i}{\partial X_A},
\end{equation}
where $x_i$ $(i=1,2,3)$ are the coordinates at time $t$ of the point whose
coordinates are $X_i$ in the undeformed reference configuration. The
Mooney-Rivlin viscoelastic solid is incompressible, which means that only
isochoric deformations are possible; in other words, the condition
\begin{equation}  \label{03b}
\det \mathbf{F}=1,
\end{equation}
must be satisfied at all times.

We assume that $\mathbf{T}^D$, the dissipative part of the stress, is given
by
\begin{equation}  \label{03c}
\mathbf{T}^D = \nu (\mathbf{L}+\mathbf{L}^\text{T}), \quad \mathbf{L} = \dot{%
\mathbf{F}}\mathbf{F}^{-1},
\end{equation}
where $\nu > 0$ is a constant. The \textit{Mooney-Rivlin viscoelastic
constitutive equation} for the Cauchy stress tensor $\mathbf{T}$ is given by
\begin{equation}  \label{1a}
\mathbf{T} = - p \mathbf{1} + C \mathbf{B} - D \mathbf{B}^{-1} + \nu (%
\mathbf{L}+\mathbf{L}^\text{T}),
\end{equation}
where $p$ is the indeterminate pressure introduced by the incompressibility
constraint (\ref{03b}), to be determined from the equations of motion and
eventual boundary conditions.

We recall that in the absence of body forces, the equations of motion are
\begin{equation}  \label{2}
\text{div } \mathbf{T} = \rho \ddot{\mathbf{x}},
\end{equation}
where $\rho$ is the constant mass density of the material.


\subsection{Transverse inhomogeneous motions in a homogeneously deformed
material}


Consider a finite static isochoric homogeneous deformation defined by
\begin{equation}  \label{3}
\mathbf{x}=\mathbf{FX},
\end{equation}
where the $F_{iA}$ are constant and $\text{det } \mathbf{F} = 1$. For the
deformation \eqref{3} both $\mathbf{B}$ and $\mathbf{B}^{-1}$ are constant
and $\mathbf{L}= \mathbf{0}$. Then the following constant Cauchy stress
tensor,
\begin{equation}  \label{3a}
\mathbf{T} = - p_0\mathbf{1} + C \mathbf{B} - D\mathbf{B}^{-1}, \quad p_0 =
\text{const.},
\end{equation}
clearly satisfies the equilibrium equations: $\text{div } \mathbf{T} =
\mathbf{0}$.

On this state of static deformation, superpose a finite motion taking the
particle at $\mathbf{x}$ to $\overline{\mathbf{x}}$, given by
\begin{equation}  \label{3b}
\overline{\mathbf{x}} = \mathbf{x} + \mathbf{a} f(\mathbf{b \cdot x}) g(%
\mathbf{n\cdot x}-vt).
\end{equation}
Here $f$ and $g$ are functions to be determined and $\mathbf{n}$, $\mathbf{b}
$, and $\mathbf{a}$ are linearly independent unit vectors, such that $%
\mathbf{b} \times \mathbf{n}\ne \mathbf{0}$, $\mathbf{b \cdot a} = \mathbf{n
\cdot a} = 0$. The motion \eqref{3b} represents a transverse inhomogeneous
plane wave, propagating with speed $v$ in the direction of $\mathbf{n}$,
linearly-polarized in the direction of $\mathbf{a}$. When the wave is not
damped, the amplitude varies in the direction of $\mathbf{b}$; when the wave
is damped, the amplitude varies in some direction in the ($\mathbf{b},
\mathbf{n}$)-plane, to be determined from the equations of motion. The case
where $f$ is a constant (homogeneous waves) has already been considered by
Hayes and Saccomandi \cite{HS1}; the case of an \textit{elastic}
Mooney-Rivlin material ($\nu = 0$ in \eqref{1a}) has been considered by
Destrade \cite{D}.

For the motion \eqref{3b}, the deformation gradient $\overline{\mathbf{F}}$
is calculated using the chain rule as
\begin{equation}  \label{3c}
\overline{\mathbf{F}} = \mathbf{\check{F} F}, \quad \mathbf{\check{F}} :=
\mathbf{1} + f^{\prime }g\mathbf{a\otimes b} + fg^{\prime }\mathbf{a\otimes n%
}.
\end{equation}
Taking the tensor product of $\mathbf{\check{F}}$ by $\mathbf{1} - f^{\prime
}g\mathbf{a\otimes b} - fg^{\prime }\mathbf{a\otimes n}$, we find the
identity tensor, and so
\begin{equation}  \label{Fminus1}
\overline{\mathbf{F}}^{-1} = \mathbf{F}^{-1}\mathbf{\check{F}}^{-1}, \quad
\mathbf{\check{F}}^{-1} = \mathbf{1} - f^{\prime }g\mathbf{a\otimes b} -
fg^{\prime }\mathbf{a\otimes n}.
\end{equation}

It follows that the left Cauchy-Green tensor $\overline{\mathbf{B}}$ and its
inverse $\overline{\mathbf{B}}^{-1}$ are given by
\begin{equation}  \label{3d}
\overline{\mathbf{B}} = \mathbf{\check{F}}\mathbf{B} \mathbf{\check{F}}^%
\text{T}, \quad \overline{\mathbf{B}}^{-1} = \mathbf{\check{F}}^{-\text{T}}%
\mathbf{B}^{-1} \mathbf{\check{F}}^{-1},
\end{equation}
where $\mathbf{\check{F}}^{-\text{T}}$ is a shorthand notation for $(\mathbf{%
\check{F}}^{-1})^{\text{T}}$. Also, the time derivative of $\overline{%
\mathbf{F}}$ is given by
\begin{equation}
\dot{\overline{\mathbf{F}}} = \mathbf{\dot{\check{F}}}\mathbf{F}, \quad
\text{where} \quad \mathbf{\dot{\check{F}}} = -v (f^{\prime}g^{\prime}%
\mathbf{a\otimes b} - fg^{\prime\prime}\mathbf{a\otimes n}).
\end{equation}
Hence, the velocity gradient $\overline{\mathbf{L}}$ is here
\begin{equation}  \label{L}
\overline{\mathbf{L}} = \dot{\overline{\mathbf{F}}} \; \overline{\mathbf{F}}%
^{-1} = \mathbf{\dot{\check{F}}} \; \mathbf{\check{F}}^{-1} =
-v[f^{\prime}g^{\prime}\mathbf{a\otimes b} + fg^{\prime\prime}\mathbf{%
a\otimes n}] = \mathbf{\dot{\check{F}}}.
\end{equation}

Now the corresponding Cauchy stress tensor $\overline{\mathbf{T}}$ (say)
takes the form,
\begin{equation}  \label{Tbar}
\overline{\mathbf{T}} = - \overline{p} \mathbf{1} + C \overline{\mathbf{B}}
- D \overline{\mathbf{B}}^{-1} + \nu (\overline{\mathbf{L}} + \overline{%
\mathbf{L}}^\text{T}),
\end{equation}
where the scalar $\overline{p}$ is to be determined from the equations of
motion.


\subsection{Equations of motion}


In order to write the equations of motion, it proves practical to introduce
the Piola-Kirchhoff stress tensor $\overline{\mathbf{P}}$ associated with
the motion \eqref{3b} and defined when the static state of pure homogeneous
finite deformation is considered to be the reference configuration. Hence,
\begin{equation}  \label{PK}
\overline{P}_{ik} := \overline{T}_{ij} \check{F}^{-1}_{kj} = - \overline{p}
\check{F}^{-1}_{ki} + C \check{F}_{ip} B_{pk} - D \overline{B}^{-1}_{ij}
\check{F}^{-1}_{kj} + \nu \dot{\check{F}}_{ij} \check{F}^{-1}_{kj} + \nu
\dot{\check{F}}_{ki}.
\end{equation}
and the equations of motion read
\begin{equation}  \label{motionPK}
\rho \partial^2 \overline{x}_i / \partial t^2 = \partial \overline{P}_{ik} /
\partial x_k.
\end{equation}

We introduce the real variables
\begin{equation}
\zeta:= \mathbf{b \cdot x}, \quad \eta:= \mathbf{n \cdot x}, \quad \text{so
that} \quad f = f(\zeta), \quad g = g(\eta - vt).
\end{equation}
Now we compute in turn the terms of \eqref{motionPK}, using the expansion %
\eqref{PK}$_2$ for $\overline{P}_{ik}$. First, the left hand-side term,
\begin{equation}  \label{term0}
\rho \partial^2 \overline{x}_i / \partial t^2 = \rho v^2 f
g^{\prime\prime}a_i.
\end{equation}

Next, the first term on the right hand-side,
\begin{equation}  \label{term1}
-\partial (\overline{p} \check{F}^{-1}_{ki}) / \partial x_k = - (\partial
\overline{p} / \partial x_k) \check{F}^{-1}_{ki} = - (\overline{p}_{,\zeta}
b_k + \overline{p}_{,\eta} n_k) \check{F}^{-1}_{ki} = - \overline{p}%
_{,\zeta} b_i - \overline{p}_{,\eta} n_i,
\end{equation}
where we used the Euler-Jacobi-Piola (EJP) identity for incompressible
materials $\partial \check{F}^{-1}_{ki}/ \partial x_k=0$ in the first
equality, and $a_k n_k = a_k b_k = 0$ in the last.

Now, the second term on the right hand-side,
\begin{multline}  \label{term2}
\partial (\check{F}_{ip} B_{pk}) / \partial x_k = (\partial \check{F}_{ip} /
\partial x_k) B_{pk} \\
= [f^{\prime\prime}g(\mathbf{b \cdot Bb}) +2f^{\prime}g^{\prime}(\mathbf{n
\cdot Bb}) +fg^{\prime\prime}(\mathbf{n \cdot Bn})]a_i.
\end{multline}

Then, the expansion of the third term on the right hand-side requires the
EJP identity,
\begin{align}  \label{term3}
\partial ( \overline{B}^{-1}_{ij} & \check{F}^{-1}_{kj})/\partial x_k =
(\partial \overline{B}^{-1}_{ij} / \partial x_k) \check{F}^{-1}_{kj} =
(\partial \overline{B}^{-1}_{ij} / \partial x_k) \delta_{kj} = \partial
\overline{B}^{-1}_{ij} / \partial x_j  \notag \\
= & -[f^{\prime\prime}g(\mathbf{a \cdot B^{-1}b}) + f^{\prime}g^{\prime}(%
\mathbf{n \cdot B^{-1}n})] b_i  \notag \\
& -[f^{\prime}g^{\prime}(\mathbf{a \cdot B^{-1}b}) + fg^{\prime\prime}(%
\mathbf{n \cdot B^{-1}n})] n_i  \notag \\
& -[f^{\prime\prime}g + 2f^{\prime}g^{\prime}(\mathbf{n \cdot b}) +
fg^{\prime\prime}]B^{-1}_{ij}a_j  \notag \\
& + [2f^{\prime}f^{\prime\prime}g^2 +
(3f^{^{\prime}2}+ff^{\prime\prime})gg^{\prime}(\mathbf{n \cdot b}) +
ff^{\prime}(g^{^{\prime}2}+gg^{\prime\prime})] (\mathbf{a \cdot} \mathbf{B}%
^{-1}\mathbf{a})b_i  \notag \\
& + [2f^2g^{\prime}g^{\prime\prime}+
ff^{\prime}(3g^{^{\prime}2}+gg^{\prime\prime})(\mathbf{n \cdot b}) +
(f^{^{\prime}2}+ff^{\prime\prime})gg^{\prime}] (\mathbf{a \cdot} \mathbf{B}%
^{-1}\mathbf{a})n_i.
\end{align}

The expansion of the fourth term on the right hand-side also involves the
EJP identity,
\begin{align}  \label{term4}
\partial (\dot{\check{F}}^{-1}_{ij}\check{F}^{-1}_{kj})/\partial x_k =
&(\partial \dot{\check{F}}^{-1}_{ij}/\partial x_k)\check{F}^{-1}_{kj}  \notag
\\
= & -v[f^{\prime\prime}g^{\prime}b_k b_j + f^{\prime}g^{\prime\prime}(n_k
b_j + b_k n_j) + fg^{\prime\prime\prime}n_k n_j] a_i \check{F}^{-1}_{kj}
\notag \\
= & -v[f^{\prime\prime}g^{\prime}+ 2f^{\prime}g^{\prime\prime}(\mathbf{n
\cdot b}) + fg^{\prime\prime\prime}] a_i.
\end{align}

Finally the fifth and last term on the right hand-side turns out to be zero:
\begin{equation}  \label{term5}
\partial \dot{\check{F}}^{-1}_{ki} / \partial x_k =
-v[f^{\prime\prime}g^{\prime}b_k a_k b_i + f^{\prime}g^{\prime\prime}n_k a_k
b_i + f^{\prime}g^{\prime\prime}b_k a_k n_i + fg^{\prime\prime\prime}n_k a_k
n_i] =0,
\end{equation}
because $a_k n_k = a_k b_k =0$.

Now we proceed to the resolution of the equations of motion \eqref{motionPK}
using \eqref{PK} and \eqref{term0}-\eqref{term5}. We treat in turn the cases
of a Mooney-Rivlin viscoelastic material ($C>0, D>0, \nu>0$), of a
neo-Hookean viscoelastic material ($C>0, D=0, \nu>0$), and of a Newtonian
viscous fluid ($C=0, D=0, \nu>0$).


\section{Mooney-Rivlin viscoelastic solid}



\subsection{Orthogonality of $\mathbf{n}$ and $\mathbf{b}$}


Destrade \cite{D} proved that finite-amplitude inhomogeneous plane waves of
evanescent sinusoidal type may propagate in a homogeneously deformed
Mooney-Rivlin \textit{elastic} material only when the planes of constant
phase are orthogonal to the planes of constant amplitude. Here we prove that
this result is not affected by the addition in \eqref{1a} of a \textit{%
viscous} part to the Cauchy stress.

Take the superposed motion $\mathbf{a} f(\mathbf{b \cdot x}) g(\mathbf{%
n\cdot x}-vt)$ in \eqref{3b} to be of evanescent sinusoidal type,
\begin{equation}
f = \alpha \text{e}^{-\omega \sigma \zeta}, \quad g = 2\beta \cos \omega
v^{-1}(\eta -vt),
\end{equation}
where the amplitudes $\alpha$, $\beta$, the frequency $\omega$, and the
attenuation factor $\sigma$ are arbitrary real scalars. Then, leaving aside
the pressure terms \eqref{term1} for the time being, we find by inspection
of \eqref{term0} and \eqref{term2}-\eqref{term5} that all the other terms
appearing in the equations of motion \eqref{motionPK} can be decomposed
along the set of linearly independent functions $\text{e}^{-\omega \sigma
\zeta} \cos \omega v^{-1}(\eta -vt)$, $\text{e}^{-\omega \sigma \zeta} \sin
\omega v^{-1}(\eta -vt)$, $\text{e}^{-2\omega \sigma \zeta} \cos 2\omega
v^{-1}(\eta -vt)$, $\text{e}^{-2\omega \sigma \zeta} \sin 2\omega
v^{-1}(\eta -vt)$, and $\text{e}^{-2\omega \sigma \zeta}$. The only terms
along this latter (time-independent) function come from the decomposition %
\eqref{term3}, where we have
\begin{multline}
2 f^{\prime}f^{\prime\prime}g^2 + ff^{\prime}g^{^{\prime}2} = - 2\alpha^2
\beta^2 \omega^3 \sigma (2\sigma^2 + v^{-2}) \text{e}^{-2\omega \sigma
\zeta} + \ldots \\
\text{and} \quad ff^{\prime}(3g^{^{\prime}2} + gg^{\prime\prime}) = -
4\alpha^2 \beta^2 \omega^3 \sigma v^{-2} \text{e}^{-2\omega \sigma \zeta} +
\ldots
\end{multline}
Here the ellipses stand for terms proportional to time-dependent functions.
It follows from the equations of motion that $\overline{p}$ must also be of
this form,
\begin{equation}
\overline{p} = - \alpha^2 \beta^2 \omega^2 p_1 \text{e}^{-2\omega \sigma
\zeta} + \ldots,
\end{equation}
where $p_1$ is constant. Then the time-independent part of the equations of
motion \eqref{motionPK} is simply
\begin{equation}
\mathbf{0} = [-p_1 + D(2\sigma^2 + v^{-2})(\mathbf{a \cdot} \mathbf{B}^{-1}
\mathbf{a})] \mathbf{b} + 2D v^{-2} (\mathbf{n \cdot b}) (\mathbf{a \cdot}
\mathbf{B}^{-1} \mathbf{a}) \mathbf{n}.
\end{equation}
Because $\mathbf{n}$ and $\mathbf{b}$ cannot be parallel for the motion %
\eqref{3b} to be inhomogeneous, we conclude that $p_1 = D(2\sigma^2 +
v^{-2})(\mathbf{a \cdot} \mathbf{B}^{-1} \mathbf{a})$ and that
\begin{equation}
\mathbf{n \cdot b} = 0.
\end{equation}
This important result, established in \cite{D} for a Mooney-Rivlin elastic
solid, still holds when the solid is viscoelastic with a constitutive
equation of the form \eqref{1a} because as we just saw, the viscous terms %
\eqref{term4} contribute to the equations of motion only with time-dependent
functions. Turning back to a general inhomogeneous motion of the form %
\eqref{3b}, this result leads us to assume that here also, $\mathbf{n}$ and $%
\mathbf{b}$ are orthogonal, so that ($\mathbf{n}, \mathbf{b}, \mathbf{a}$)
is an orthonormal basis. Accordingly, some simplifications occur in the
equations of motion.

We introduce the notation
\begin{align}  \label{VnVb}
& \rho v_n^2 := C (\mathbf{n \cdot B n}) + D (\mathbf{a}\cdot \mathbf{B}^{-1}%
\mathbf{a})>0,  \notag \\
& \rho v_b^2 := C (\mathbf{b \cdot B b}) + D (\mathbf{a}\cdot \mathbf{B}^{-1}%
\mathbf{a})>0,  \notag \\
& \rho b := C (\mathbf{n \cdot B b}).
\end{align}
Here $v_n$ ($v_b$) is the speed of an homogeneous finite-amplitude plane
wave propagating in the direction of $\mathbf{n}$ ($\mathbf{b}$) and
polarized in the direction of $\mathbf{a}$, as proved by Boulanger and Hayes
\cite{BoHa95}. Then the projection of the equation of motion \eqref{motionPK}
along $\mathbf{a}$ is written in compact form as
\begin{equation}  \label{balance}
\rho v_b^2 f^{\prime\prime}g + 2 \rho b f^{\prime}g^{\prime} - \nu
v(fg^{\prime\prime\prime} + f^{\prime\prime}g^{\prime}) + \rho (v_n^2 -
v^2)fg^{\prime\prime}= 0.
\end{equation}
Finally, the projections of the equations of motion \eqref{motionPK} along $%
\mathbf{n} $ and along $\mathbf{b}$ yield expressions for the derivatives $%
\overline{p}_{,\eta}$ and $\overline{p}_{,\zeta}$; writing $\overline{p}%
_{,\eta \zeta} = \overline{p}_{,\eta \zeta}$ gives \cite{D},
\begin{multline}  \label{compatibility}
(f^{\prime\prime\prime}g + f^{\prime}g^{\prime\prime}) (\mathbf{a \cdot
B^{-1}n}) - (f^{\prime\prime}g^{\prime}+ fg^{\prime\prime\prime}) (\mathbf{a
\cdot B^{-1}b}) \\
+ [(f^{\prime}f^{\prime\prime}- ff^{\prime\prime\prime})gg^{\prime}-
(g^{\prime}g^{\prime\prime}- gg^{\prime\prime\prime})ff^{\prime}](\mathbf{%
a\cdot B^{-1}a}) = 0.
\end{multline}
Hence the propagation of finite-amplitude, linearly polarized, transverse,
plane waves in a deformed Mooney-Rivlin viscoelastic solid is governed by
two equations: the ``balance equation'' \eqref{balance} and the
``compatibility equation'' \eqref{compatibility}, to be solved
simultaneously.


\subsection{Derivation of the solutions}


First divide \eqref{balance} by $f(\zeta)g^{\prime}(\eta - vt)$ to get
\begin{equation}  \label{balance2}
\rho v_b^2 \frac{f^{\prime\prime}}{f} \frac{g}{g^{\prime}} + 2 \rho b \frac{%
f^{\prime}}{f} - \nu v (\frac{g^{\prime\prime\prime}}{g^{\prime}} + \frac{%
f^{\prime\prime}}{f}) + \rho (v_n^2 - v^2) \frac{g^{\prime\prime}}{g^{\prime}%
} = 0.
\end{equation}
Then differentiate first with respect to the argument of $f$, next with
respect to the argument of $g$, to obtain
\begin{equation}
(f^{\prime\prime}/f)^{\prime}(g/g^{\prime})^{\prime}= 0,
\end{equation}
so that either $f^{\prime\prime}/f$ or $g/g^{\prime}$ is a constant. The
following cases cover all possibilities,
\begin{equation}
(i) \: f^{\prime\prime}/f = - k_1^2, \quad (ii) \: f^{\prime\prime}/ f =
k_1^2, \quad (iii) \: g^{\prime}/g = k_2.
\end{equation}

Without loss of generality, we pick $k_1>0$ in Case \textit{(i)} and in Case
\textit{(ii)}; in Case \textit{(iii)} we take $k_2 > 0$ to avoid solutions
which grow exponentially with time $t$ (note that the forthcoming analysis 
can also be conducted to accomodate solutions which do grow exponentially
with time). We treat each case in turn, keeping
in mind that $f$ must be trigonometric when $g$ is exponential and \textit{%
vice-versa}, for the motion to be inhomogeneous \cite{D}.

\vspace{11pt}

\textit{Case (i)}: $f(\zeta) = A \cos k_1 \zeta + B \sin k_1 \zeta$.

\noindent Substitution of this form of solution into the balance equation %
\eqref{balance2} and differentiation with respect to $\zeta$ yields $\rho b
(f^{\prime}/f)^{\prime}= 0$ and so,

\begin{equation}  \label{nBb=0}
\mathbf{n \cdot B b} = 0.
\end{equation}
This condition means that the unit vectors $\mathbf{n}$ and $\mathbf{b}$
must be conjugate with respect to the central elliptical section of the $%
\mathbf{B}$-ellipsoid, $\mathbf{x \cdot B x} = 1$, by the plane orthogonal
to $\mathbf{a}$. Being orthogonal, they must be along the principal axes of
this ellipse.

Then the balance equation \eqref{balance} (with $b = 0$) gives a third order
linear differential equation for $g$,
\begin{equation}  \label{(i)}
\nu v g^{\prime\prime\prime}- \rho(v_n^2 - v^2) g^{\prime\prime}- k_1^2 \nu
v g^{\prime}+ k_1^2 \rho v_b^2 g = 0.
\end{equation}
On the other hand, the compatibility equation \eqref{compatibility} written
as an identity for $\cos k_1 \zeta$, $\sin k_1 \zeta$, $\cos 2k_1 \zeta$,
and $\sin 2k_1 \zeta$ yields
\begin{equation}
g^{\prime}g^{\prime\prime}- gg^{\prime\prime\prime}= 0, \quad
(g^{\prime\prime}- k_1^2 g)(\mathbf{a} \cdot \mathbf{B}^{-1} \mathbf{n}) =0,
\quad (g^{\prime\prime\prime}- k_1^2 g^{\prime})(\mathbf{a} \cdot \mathbf{B}%
^{-1} \mathbf{b}) = 0.
\end{equation}
These three equations are satisfied simultaneously either when \textit{(a)} $%
g^{\prime\prime}= k_1^2 g$ or when \textit{(b)} $g^{\prime\prime}/g =$%
const., $\mathbf{a} \cdot \mathbf{B}^{-1} \mathbf{n} = \mathbf{a} \cdot
\mathbf{B}^{-1} \mathbf{b} = 0$.

In \textit{Case (ia)}, $g$ is an exponential function. Discarding solutions
which blow up with time, we find
\begin{equation}  \label{solution1}
g(\eta-vt)= \text{e}^{k_1 (\eta-vt)}.
\end{equation}
Substitution into the third-order differential equation \eqref{(i)} fixes $v$
as
\begin{equation}  \label{v}
\rho v^2 = \rho (v_n^2 - v_b^2) = C[(\mathbf{n \cdot Bn}) - (\mathbf{b \cdot
Bb})].
\end{equation}
The quantity $v$ is real when $\mathbf{n}$ and $\mathbf{b}$ are in the
respective directions of the minor and major axes of the elliptical section
of the $\mathbf{B}$-ellipsoid by the plane orthogonal to $\mathbf{a}$.

In \textit{Case (ib)}, $g$ is of the form
\begin{equation}  \label{principal1}
g(\eta-vt)= \text{e}^{k_2 (\eta-vt)},
\end{equation}
where $k_2>0$ to avoid blowing-up solutions. The conditions on $\mathbf{n}$,
$\mathbf{b}$, $\mathbf{a}$, together with \eqref{nBb=0}, imply that these
unit vectors are along \textit{principal directions} of the $\mathbf{B}$%
-ellipsoid. Substitution into the third-order differential equation %
\eqref{(i)} relates $k_1$ to $k_2$ through
\begin{equation}  \label{k1}
k_1^2 = \frac{\rho(v^2_n - v^2) - k_2 \nu v}{\rho v^2_b - k_2 \nu v} k_2^2.
\end{equation}
Recall that $k_1$ is assumed real so that, in order to construct a solution
here, we may pick any values for $k_2>0$ and for $v>0$, as long as $%
\rho(v^2_n - v^2) - k_2 \nu v$ and $\rho v^2_b - k_2 \nu v$ are of the same
sign, where $v_n$, $v_b$ are given by \eqref{VnVb} with $\mathbf{n}$, $%
\mathbf{b}$, $\mathbf{a}$ along principal directions of the $\mathbf{B}$%
-ellipsoid. Then $k_1$ is given by \eqref{k1}.

\vspace{11pt}

\textit{Case (ii)}: $f(\zeta) = A \cosh k_1 \zeta + B \sinh k_1 \zeta$.

\noindent Substitution of this form of solution into the balance equation %
\eqref{balance2} and differentiation with respect to $\zeta$ yields the
condition \eqref{nBb=0} as in \textit{Case (i)}.

Then the balance equation \eqref{balance} (with $b = 0$) gives a third order
linear differential equation for $g$,
\begin{equation}  \label{(ii)}
\nu v g^{\prime\prime\prime}- \rho(v_n^2 - v^2) g^{\prime\prime}+ k_1^2 \nu
v g^{\prime}- k_1^2 \rho v_b^2 g = 0.
\end{equation}
Also, the compatibility equation \eqref{compatibility} written as an
identity for $\cosh k_1 \zeta$, $\sinh k_1 \zeta$, $\cosh 2k_1 \zeta$, and $%
\sinh 2k_1 \zeta$ yields
\begin{equation}
g^{\prime}g^{\prime\prime}- gg^{\prime\prime\prime}= 0, \quad
(g^{\prime\prime}+ k_1^2 g)(\mathbf{a} \cdot \mathbf{B}^{-1} \mathbf{n}) =0,
\quad (g^{\prime\prime\prime}+ k_1^2 g^{\prime})(\mathbf{a} \cdot \mathbf{B}%
^{-1}\mathbf{b}) = 0.
\end{equation}
These three equations are satisfied simultaneously either when \textit{(a)} $%
g^{\prime\prime}= -k_1^2 g$ or when \textit{(b)} $g^{\prime\prime}/g =$%
const., $\mathbf{a} \cdot \mathbf{B}^{-1} \mathbf{n} = \mathbf{a} \cdot
\mathbf{B}^{-1} \mathbf{b} = 0$.

In \textit{Case (iia)}, $g$ is of the form
\begin{equation}  \label{solution2}
g(\eta-vt)= d_1 \cos k_1 (\eta-vt) + d_2 \sin k_1 (\eta -vt),
\end{equation}
and substitution into the third-order differential equation \eqref{(ii)}
fixes $v$ again as \eqref{v}, $v^2 = v_n^2 - v_b^2$.

In \textit{Case (iib)}, $g$ is of the form $g = d_1 \cos k_2 (\eta-vt) + d_2
\sin k_2 (\eta -vt)$, where $k_2$ is arbitrary. The conditions on $\mathbf{n}
$, $\mathbf{b}$, $\mathbf{a}$, together with \eqref{nBb=0} imply that these
unit vectors are along principal directions. Then substitution of $g$ into
the third-order differential equation \eqref{(ii)} gives
\begin{equation}
\nu v (k_1^2 - k_2^2) g^{\prime}- \rho [k_1^2 v_b^2 - k_2^2 (v_n^2 - v^2)] g
= 0.
\end{equation}
This equation is satisfied for $g$ trigonometric and $\nu \ne 0$ only when
the respective coefficients of $g$ and $g^{\prime}$ are zero, conditions
which lead back to a solution of the form \eqref{solution2} and $v$ uniquely
determined by \eqref{v}. The existence of these `special principal motions'
is therefore more limited than when the solid is purely elastic \cite{D}
because there they may propagate at an arbitrary speed $v$ within the
interval $[0,v_n]$.

\vspace{11pt}

\textit{Case (iii)}: $g(\eta - vt) = \text{e}^{k_2 (\eta - vt)}$.

Substituting this form of solution into the balance equation \eqref{balance}%
, we obtain a second-order linear differential equation for $f$,
\begin{equation}  \label{(iii)b}
(\rho v_b^2 - k_2 \nu v) f^{\prime\prime}+ 2 k_2 \rho b f^{\prime}+ k_2^2
[\rho (v_n^2 - v^2) - k_2 \nu v] f = 0.
\end{equation}
Also, the compatibility equation \eqref{compatibility} reduces to
\begin{multline}
(f^{\prime\prime\prime}+ k_2^2 f^{\prime})(\mathbf{a} \cdot \mathbf{B}^{-1}%
\mathbf{n}) - k_2(f^{\prime\prime}+ k_2^2 f)(\mathbf{a} \cdot \mathbf{B}%
^{-1} \mathbf{b}) \\
+ k_2 (f^{\prime}f^{\prime\prime}- ff^{\prime\prime\prime}) (\mathbf{a}
\cdot \mathbf{B}^{-1} \mathbf{a})\text{e}^{k_2 (\eta-vt)} = 0.
\end{multline}
Differentiating this equation with respect to $(\eta - vt)$, we find that $%
f^{\prime\prime}/f =$ const. This constant must be negative so that the
combination of $f$ and $g$ represents an inhomogeneous motion \cite{D}. But
if $f$ is trigonometric, then the second order differential equation %
\eqref{(iii)b} can be satisfied only when $\rho b = C(\mathbf{n} \cdot
\mathbf{B} \mathbf{b}) = 0$, which leads us back to \textit{Case (i)} and
the solutions \eqref{solution1}-\eqref{v} (non-principal motions) and %
\eqref{principal1}-\eqref{k1} (principal motions).


\subsection{Summary of results and comparison with the purely elastic case}


For a deformed \textit{elastic} Mooney-Rivlin material, Destrade \cite{D}
proved that only two types of finite-amplitude inhomogeneous motions are
possible: either $f$ is trigonometric and $g$ is exponential,
\begin{equation}  \label{D1}
f(\zeta)=a_1 \cos k_1 \zeta +a_2 \sin k_1 \zeta, \quad g(\eta-vt)= d_1 \text{%
e}^{k_2 (\eta-vt)},
\end{equation}
or $f$ is hyperbolic and $g$ is trigonometric,
\begin{equation}  \label{D2}
f(\zeta)=a_1 \cosh k_1 \zeta +a_2 \sinh k_1 \zeta, \quad g(\eta-vt)= d_1
\cos k_2 (\eta-vt) + d_2 \sin k_2 (\eta -vt),
\end{equation}
where $a_1, a_2, d_1, d_2$ are constants and $k_2$ is arbitrary.

When the vectors $(\mathbf{a}, \mathbf{b}, \mathbf{n})$ are not aligned with
the principal axes of deformation, then the quantities $k_1$ and $v$ are
determined by
\begin{equation}  \label{D3}
k_1 = k_2, \quad \rho v^2 = \rho (v_n^2 - v_b^2) = C[(\mathbf{n \cdot Bn}) -
(\mathbf{b \cdot Bb})],
\end{equation}
and the unit vectors $\mathbf{n}$ and $\mathbf{b}$ must be conjugate with
respect to the $\mathbf{B}$-ellipsoid, $\mathbf{x \cdot B x} = 1$, that is
the condition \eqref{nBb=0} must be satisfied.

When the vectors $(\mathbf{a}, \mathbf{b}, \mathbf{n})$ are aligned with the
principal axes of deformation (`special principal motions'), then the
quantity $v$ may take \textit{any} value within the interval $[0,v_n]$ and $%
k_1$ is given by
\begin{equation}  \label{k1elastic}
k_1^2 = \frac{v_n^2 - v^2}{v_b^2} k_2^2.
\end{equation}

Here we saw that the corresponding results for a deformed \textit{%
viscoelastic} Mooney-Rivlin material are essentially the same as above, with
the following difference. When the three vectors $\mathbf{a}, \mathbf{b},
\mathbf{n}$ are not aligned with the principal axes of deformation, then the
solution is indeed of the form \eqref{D1}-\eqref{D3} with the condition %
\eqref{nBb=0}. However, when all three vectors $\mathbf{a}, \mathbf{b},
\mathbf{n}$ are aligned with the principal axes of deformation, then the
solution is of the form \eqref{D1} or \eqref{D2} but: when $f$ is
trigonometric and $g$ is exponential as in \eqref{D1}, then $k_1$ is given
by \eqref{k1}, and $v$, $k_2$ are arbitrary as long as $k_1^2$ is positive;
when $f$ is hyperbolic and $g$ is trigonometric as in \eqref{D2}, then $v$
are determined by \eqref{v} that is, $v$ is fixed and no longer arbitrary as
in the elastic case.

Now we see that when $D=0$, the visco-elastic solid allows more solutions
than its elastic counterpart.


\section{Neo-Hookean viscoelastic solid}


When $D = 0$ ($C \neq 0$) the strain-energy density \eqref{01} reduces to
the well-known neo-Hookean form. A great variety of exact solutions is
uncovered, in particular solutions in the form of finite-amplitude damped
inhomogeneous waves. Note that this problem was recently studied in the
\textit{elastic} compressible case by Rodrigues Ferreira and Boulanger \cite
{RB}.

We start with the equations of motion \eqref{motionPK} written at $D=0$. The
unit vectors $\mathbf{b}$ and $\mathbf{n}$ are not necessarily orthogonal
but the triad ($\mathbf{a}, \mathbf{b}, \mathbf{n}$) is composed of linearly
independent vectors, so that the coefficients along each vector in the
equations of motion are all zero. The coefficients along $\mathbf{b}$
and $\mathbf{n}$ are respectively,
\begin{equation}
0 = -\overline{p}_{,\zeta}, \quad 0 = -\overline{p}_{,\eta},
\end{equation}
from which we deduce that $\overline{p}=$const. satisfies this part of the
equations of motion. The other part, along $\mathbf{a}$, is independent of $%
\overline{p}$ and reads
\begin{equation}  \label{eq2}
\rho (v^2 - v^2_n) fg^{\prime\prime}- \rho v^2_b f^{\prime\prime}g - 2 \rho
b f^{\prime}g^{\prime}+ \nu v [ f^{\prime\prime}g^{\prime}+ 2
f^{\prime}g^{\prime\prime}(\mathbf{n\cdot b}) + fg^{\prime\prime\prime}] = 0,
\end{equation}
where the quantities
\begin{equation}
\rho v^2_n = C \mathbf{n \cdot Bn}>0, \quad \rho v^2_b = C \mathbf{b \cdot Bb%
}>0, \quad \rho b = C \mathbf{n \cdot Bb},
\end{equation}
were used.

Now divide \eqref{eq2} by $f(\zeta )g^{\prime }(\eta -vt)$ and
differentiate, first with respect to $\zeta $ and then with respect to $%
(\eta -vt)$, to obtain the necessary condition
\begin{equation}
\rho v_{b}^{2}\left( \frac{f^{\prime\prime}}{f}\right)^{\prime}\left(\frac{g%
}{g^{\prime}}\right)^{\prime}=2\nu v\left( \mathbf{n\cdot b}\right) \left(
\frac{f^{\prime}}{f}\right)^{\prime}\left( \frac{g^{\prime\prime}}{g^{\prime}%
}\right) ^{\prime }.  \label{eq3}
\end{equation}
We separate the functions of different variables in this equation to
conclude that either
\begin{equation}
\left( \frac{f^{\prime}}{f}\right)^{\prime}\left( \frac{g}{g^{\prime}}%
\right)^{\prime}=0, \quad \text{or} \quad \rho v_b^2 \frac{%
(f^{\prime\prime}/f)^{\prime}}{(f^{\prime}/f)^{\prime}} = 2 \nu v (\mathbf{%
n\cdot b}) \frac{(g^{\prime\prime}/g^{\prime})^{\prime}}{(g/g^{\prime})^{%
\prime}} = \text{const.}
\end{equation}
The following cases cover all possibilities,
\begin{equation}
(i) \: f^{\prime}= k_1 f, \quad (ii) \: g^{\prime}= k_2 g,
\end{equation}
and \textit{(iii)} $\rho v_b^2 f^{\prime\prime}- k_1 f^{\prime}- k_3 f =0$, $%
2 \nu v (\mathbf{n \cdot b}) g^{\prime\prime}- k_2 g^{\prime}- k_1 g =0$,
where $k_1, k_2, k_3$ are constants. In fact, using expressions for $%
f^{\prime\prime}$, $g^{\prime\prime}$, and $g^{\prime\prime\prime}$ derived
from $(iii)$, and substituting into \eqref{eq2}, it is straightforward to
check that Case $(iii)$ leads to Cases $(i)$ or $(ii)$.

In Case $(i)$ we have
\begin{align}  \label{eq8}
& f(\zeta ) = \exp (k_1 \zeta),  \notag \\
& g^{\prime\prime\prime}+ [\frac{\rho }{\nu v}(v^2 - v_n^2) + 2 k_1(\mathbf{%
n\cdot b})]g^{\prime\prime}+ k_1(k_1 - 2\frac{\rho }{\nu v} b)g^{\prime}-
\rho \frac{v_b^2}{\nu v}k_1^2 g = 0.
\end{align}

In Case $(ii)$ we have
\begin{align}  \label{eq10}
& g(\eta -vt) = \exp [k_2(\eta -vt)],  \notag \\
& (\rho v_b^2 - k_1\nu v)f^{\prime\prime}+ 2k_2[\rho b - \nu v k_2(\mathbf{%
n\cdot b})]f^{\prime}- k_2^2 [\rho (v^2 - v_n^2) + \nu v k_2]f = 0.
\end{align}

Now consider \eqref{eq8} in greater detail. The characteristic equation
associated with the third-order differential equation for $g$ is a cubic.
Depending on the values of the coefficients (that is, on the choices for $v$
and $k_1$), the cubic has either only real roots or one real root and two
complex conjugate roots. Real roots lead to an homogeneous motion and must
be discarded. Pure imaginary roots lead to inhomogeneous plane waves with
attenuation in the direction of $\mathbf{b}$ and propagation in the
direction of $\mathbf{n}$. Non-pure-imaginary complex roots lead to
inhomogeneous \textit{damped} plane waves with attenuation in the plane of $%
\mathbf{b}$ and $\mathbf{n}$, exponential damping with time, and propagation
in the direction of $\mathbf{n}$. Because $v$ or $k_1$ may be prescribed a
priori, there is an infinity of such solutions, of which we present one
explicitly.

We choose $v$ such that the coefficient of $g^{\prime \prime }$ in %
\eqref{eq8}$_{2}$ is zero, that is we fix $v$ as
\begin{equation}
\rho v=\rho v_{o}:=k_{1}\nu (\mathbf{n\cdot b})+\sqrt{[k_{1}\nu (\mathbf{%
n\cdot b})]^{2}+(\rho v_{n})^{2}}.  \label{v0}
\end{equation}
Then \eqref{eq8} reduces to
\begin{equation}
g^{\prime \prime \prime }+\beta _{o}g^{\prime }-\gamma _{o}g=0,\quad \beta
_{o}=k_{1}(k_{1}-2\frac{\rho b}{\nu v_{o}}),\quad \gamma _{o}=k_{1}^{2}\frac{%
\rho v_{b}^{2}}{\nu v_{o}}.  \label{alpha=0}
\end{equation}
Now we seek a solution in the form
\begin{equation}
g(\eta -v_{o}t)=\text{e}^{\lambda _{o}(\eta -v_{o}t)}\cos \omega _{o}(\eta
-v_{o}t),  \label{g}
\end{equation}
where $\lambda _{o}$ and $\omega _{o}$ are real constants determined as
follows. Substituting \eqref{g} into \eqref{alpha=0}, we find in turn that $%
\omega _{o}$ and $\lambda _{o}$ satisfy
\begin{equation} 
\omega _{o}^{2}=3\lambda _{o}^{2}+\beta _{o},\quad 8\lambda _{o}^{3}+2\beta
_{o}\lambda _{o}+\gamma _{o}=0.  \label{omega-lambda}
\end{equation}
The real root of the cubic in $\lambda _{o}$ is
\begin{equation}
\lambda _{o}=\textstyle{\frac{1}{12}}[12\sqrt{12\beta _{o}^{3}+81\gamma _{o}}%
-108\gamma _{o}]^{\textstyle{\frac{1}{3}}}-\beta _{o}[12\sqrt{12\beta
_{o}^{3}+81\gamma _{o}}-108\gamma _{o}]^{-\textstyle{\frac{1}{3}}}.
\label{lambda0}
\end{equation}
That $\omega_{o}$ is real is ensured for instance when $\beta _{o}>0$ 
(see Eq.~\eqref{omega-lambda}$_1$). 
A few lines of calculation show that a sufficient condition for $\beta _{o}$
to be positive is
\begin{equation}
k_{1}^{2}>\frac{(2\rho b)^{2}}{\nu ^{2}[v_{n}^{2}+4b(\mathbf{n\cdot b})]}=%
\frac{4\rho C(\mathbf{n\cdot Bb})^{2}}{\nu ^{2}[(\mathbf{n\cdot Bn})+4(%
\mathbf{n\cdot Bb})(\mathbf{n\cdot b})]}.
\end{equation}
Thus for any \textit{arbitrary} value of $k_{1}$ satisfying this inequality,
the following finite-amplitude damped inhomogeneous plane wave may propagate
in a deformed neo-Hookean viscoelastic body,
\begin{equation}
\overline{\mathbf{x}}=\mathbf{x}+\alpha \mathbf{a}\text{e}^{(k_{1}\mathbf{b}%
+\lambda _{o}\mathbf{n})\mathbf{\cdot x}-\lambda _{o}v_{o}t}\cos \omega _{o}(%
\mathbf{n\cdot x}-v_{o}t),
\end{equation}
where $\mathbf{x}$ is given by \eqref{3}, $\alpha $ is arbitrary, the
orientation of the triad of unit vectors $\mathbf{a,b,n}$ ($\mathbf{b}$ and $%
\mathbf{n}$ orthogonal to $\mathbf{a}$) is arbitrary, and $\lambda _{o}$, $%
\omega _{o}$, and $v_{o}$ are given by \eqref{lambda0}, \eqref{omega-lambda}$%
_{1}$, and \eqref{v0}, respectively. The wave propagates in the direction of
$\mathbf{n}$, is exponentially damped with time, and is attenuated in the
direction of $k_{1}\mathbf{b}+\lambda _{0}\mathbf{n}$.

To conclude this Section, we point out that inspection of (\ref{eq3}) makes it clear 
that a neo-Hookean viscoelastic constitutive equation yields more inhomogeneous 
waves than a neo-Hookean elastic constitutive equation, where $\nu = 0$.


\section{Newtonian viscous fluids}


With $C=D=0$, equation \eqref{1a} is the constitutive relation defining an
incompressible Newtonian viscous fluid of viscosity $\nu $, and \eqref{2},
or their specialization \eqref{eq2} at $C=0$, are the (Lagrangian)
Navier-Stokes equations in the absence of body forces for the finite
amplitude inhomogeneous plane wave \eqref{3b}, that is $f^{\prime \prime
}g^{\prime }+[2(\mathbf{n\cdot b})f^{\prime }+(\rho v/\nu )f]g^{\prime
\prime }+fg^{\prime \prime \prime }=0$. Integrating with respect to the
argument of $g$ and taking the constant of integration to be zero for
simplicity, we obtain
\begin{equation}
f^{\prime \prime }g+[2(\mathbf{n\cdot b})f^{\prime }+\frac{\rho v}{\nu }%
f]g^{\prime }+fg^{\prime \prime }=0.  \label{ns}
\end{equation}

With methods similar to those presented in the two previous sections, it is
a straightforward matter to show that either $f$ or $g$ are pure exponential
functions that is, either
\begin{equation}
f(\zeta )=\exp (k_{1}\zeta ),\quad \text{and}\quad g^{\prime \prime
}+[2k_{1}(\mathbf{n\cdot b})+\frac{\rho v}{\nu }]g^{\prime }+k_{1}^{2}g=0,
\label{ns1}
\end{equation}
or
\begin{equation}
g(\eta -vt)=\exp [k_{2}(\eta -vt)],\quad \text{and}\quad f^{\prime \prime
}+2k_{2}(\mathbf{n\cdot b})f^{\prime }+k_{2}(\frac{\rho v}{\nu }+k_{2})f=0.
\label{ns2}
\end{equation}
where $k_{1}$, $k_{2}$ are arbitrary constants. The latter case represents a
standing damped wave, whatever the form of $f$ may be. The former is a
travelling inhomogeneous damped wave as long as the characteristic equation
associated with the second-order differential equation for $g$ has complex
roots. This condition reads
\begin{equation}
0<\rho v<2k_{1}\nu \lbrack 1-(\mathbf{n\cdot b})]=:\rho v_{0}.
\end{equation}
Hence the constants $k_{1}$ and $v$ are \textit{arbitrary} as long as $k_{1}$
is positive and $v$ is less than $v_{0}$ defined above. The following
displacement field $\mathbf{u}$ is solution to the Navier-Stokes equations
for any orientation of the pair of unit vectors $\mathbf{b,n}$ in the plane $%
\mathbf{a\cdot x}=0$ (which may be taken without loss of generality as $z=0$
say, because the fluid is isotropic),
\begin{align}
& u_{1}  =  0, 
\quad u_{2}  =  0, 
\nonumber   \\
& u_{3}  =  \text{e}^{k_{1}[\mathbf{b\cdot x}+\sqrt{1-\kappa ^{2}}(\mathbf{%
n\cdot x}-vt)]}[ A\cos k_{1}\kappa (\mathbf{n\cdot x}-vt) 
 B\sin k_{1}\kappa (\mathbf{n\cdot x}-vt)],
\end{align}
where $A$, $B$ are constants and $\kappa $ is defined by
\begin{equation}
\kappa :=\sqrt{1-[(\mathbf{n\cdot b})+\frac{\rho v}{2k_{1}\nu }]^{2}}.
\end{equation}
This finite-amplitude wave propagates in the direction of $\mathbf{n}$ with
speed $v$ and wave number $k_{1}\kappa $, is damped exponentially with
respect to time, and is attenuated in the direction of $\mathbf{b}+\sqrt{%
1-\kappa ^{2}}\mathbf{n}$. The angle $\theta $ (say) between the normal to
the planes of constant phase and the normal to the planes of constant
amplitude is given by
\begin{equation}
\cot \theta =\mathbf{n\cdot b}+\sqrt{1-\kappa ^{2}}=2(\mathbf{n\cdot b})+%
\frac{\rho v}{k_{1}\nu }.
\end{equation}

Note that Boulanger, Hayes, and Rajagopal \cite{BoHR91} also investigated
the possibility of linearly-polarized, damped, inhomogeneous plane wave
solutions to the Navier-Stokes equations. However, they prescribed \textit{a
priori} the form of the solution. The result presented above allows for a
greater variety of solutions, in the same manner that starting from a
finite-amplitude wave in the form \eqref{3b} where $f$ and $g$ are unknown
functions, rather than prescribed as exponential and sinusoidal functions,
allowed Rodrigues Ferreira and Boulanger \cite{RB} to find more solutions
than Destrade \cite{Dest99} in the elastic neo-Hookean case.

We also note that the solutions found here
for the Navier-Stokes equations are a generalization of the
celebrated \emph{Kelvin modes} \cite{Kelvin}. These are
disturbances in the form of homogeneous plane waves of particular
interest in the stability study of basic flows characterized by
spatially uniform shearing rates, i.e. self-equilibrated
homogeneous motions \cite{Truesdell}. This is because Craik and
Criminale \cite{CC} have shown that the superimposition of a
single Kelvin mode to the above mentioned basic flows is an exact
solution of the full Navier-Stokes equations. 
It would be interesting to investigate if the superimposition of our solutions
to homogeneous self-equilibrated motions will give some new exact
solutions for the Navier-Stokes equations, but this investigation
is outside the framework of the present paper.

\end{document}